\documentclass[aps,prl,twocolumn,showpacs,amsmath,amssymb,superscriptaddress]{revtex4}
\usepackage{graphicx}
\usepackage{amssymb}
\usepackage{natbib}
\usepackage{calc}
\usepackage{SIunits}
\usepackage{textcomp}

\begin{document}

\title{Neutron Spin Resonance in the 112-Type Iron-Based Superconductor}

\author{Tao Xie}
\affiliation{Beijing National Laboratory for Condensed Matter
Physics, Institute of Physics, Chinese Academy of Sciences, Beijing
100190, China}
\affiliation{University of Chinese Academy of Sciences, Beijing 100049, China}
\author{Dongliang Gong}
\affiliation{Beijing National Laboratory for Condensed Matter
Physics, Institute of Physics, Chinese Academy of Sciences, Beijing
100190, China}
\affiliation{University of Chinese Academy of Sciences, Beijing 100049, China}
\author{Haranath Ghosh}
\affiliation{Homi Bhabha National Institute, Anushakti Nagar, Mumbai 400094, India}
\affiliation{Human Resources development section, Raja Ramanna Centre for Advanced Technology, Indore - 452013, India}
\author{Abyay Ghosh}
\affiliation{Homi Bhabha National Institute, Anushakti Nagar, Mumbai 400094, India}
\affiliation{Human Resources development section, Raja Ramanna Centre for Advanced Technology, Indore - 452013, India}
\author{Minoru Soda}
\affiliation{The Institute for Solid State Physics, The University of Tokyo, Chiba 277-8581, Japan}
\author{Takatsugu Masuda}
\affiliation{The Institute for Solid State Physics, The University of Tokyo, Chiba 277-8581, Japan}
\author{Shinichi Itoh}
\affiliation{Institute of Materials Structure Science, High Energy Accelerator Research Organization, Tsukuba, Ibaraki 305-0801, Japan}
\author{Fr\'{e}d\'{e}ric Bourdarot}
\affiliation{Universit\'{e} Grenoble Alpes, CEA, INAC, MEM MDN, F-38000 Grenoble, France}
\author{Louis-Pierre Regnault}
\affiliation{Intitut Laue Langevin, 71 avenue des Martyrs, CS 20156, 38042 Grenoble Cedex, France}
\author{Sergey Danilkin}
\affiliation{Australian Centre for Neutron Scattering, Australian Nuclear Science and
Technology Organization, Lucas Heights NSW-2234, Australia}
\author{Shiliang Li}
\affiliation{Beijing National Laboratory for Condensed Matter
Physics, Institute of Physics, Chinese Academy of Sciences, Beijing
100190, China}
\affiliation{University of Chinese Academy of Sciences, Beijing 100049, China}
\affiliation{Collaborative Innovation Center of Quantum Matter, Beijing 100190, China}
\author{Huiqian Luo}
\email{hqluo@iphy.ac.cn}
\affiliation{Beijing National Laboratory for Condensed Matter
Physics, Institute of Physics, Chinese Academy of Sciences, Beijing 100190, China}

\date{\today}
\pacs{74.70.Xa, 75.30.Gw, 78.70.Nx, 75.40.Gb}

\begin{abstract}

We use inelastic neutron scattering to study the low-energy spin excitations of the 112-type iron pnictide Ca$_{0.82}$La$_{0.18}$Fe$_{0.96}$Ni$_{0.04}$As$_{2}$ with bulk superconductivity below $T_c=22$ K.
A two-dimensional spin resonance mode is found around $E=$ 11 meV, where the resonance energy is almost temperature independent and linearly scales with $T_c$ along with other iron-based superconductors. Polarized neutron analysis reveals the resonance is nearly isotropic in spin space without any $L$ modulations. Because of the unique monoclinic structure with additional zigzag arsenic chains, the As $4p$ orbitals contribute to a three-dimensional hole pocket around the $\Gamma$ point and an extra electron pocket at the $X$ point. Our results suggest that the energy and momentum distribution of the spin resonance does not directly respond to the $k_z$ dependence of the fermiology, and the spin resonance intrinsically is a spin-1 mode from singlet-triplet excitations of the Cooper pairs in the case of weak spin-orbital coupling.

\end{abstract}

\maketitle

In unconventional superconductors such as copper oxides, heavy fermions, and iron-based superconductors, the neutron spin resonance is a crucial evidence for spin fluctuation mediated superconductivity in the proximity of an antiferromagnetic (AF) instability.  On cooling below the superconducting transition temperature $T_c$,  the intensity of spin excitations around a particular energy (the so called resonance energy $E_R$) behaves like a superconducting order parameter \cite{rossat-mignod, nksato, wilson, christianson, pdai2015}. Theoretically, the spin resonance mode is generally believed to be a spin-1 exciton from the singlet-triplet excitations of the electron Cooper pairs \cite{eschrig,lipscombe,slli2010,gyu2009}. However, such a picture is still not well established yet in the iron-based superconductors, although the spin resonance has been observed in many superconducting iron pnictides and iron chalcogenides \cite{christianson,pdai2015,inosov2016,lumsden,chi,qiu,inosov2010,mook,zhang2011,ishikado,jzhao2013,chlee,zhang2013,qureshi2012,wakimoto,jtpark2011,qswang,qswang2,mwma,kiida,xie2018,msato2011,kikeuchi2014,chlee2016,masurmach2015}. The spin resonance is argued to arise from the sign-reversed ($s\pm$) quasiparticle excitations between different Fermi surfaces in these multiband systems \cite{mazin2008,kuroki2008,ding2008}. Such a mechanism should yield a sharp resonant peak with an energy below the total superconducting gaps summed on the nesting bands \cite{gyu2009,mazin2008,xie2018}, while some exceptions in particular compounds with a broad enhancement of intensity and $E_R>2\Delta$ are argued to be the sign-preserved ($s_{++}$) superconducting state \cite{mwma,msato2011,qswang2,masurmach2015,kikeuchi2014,chlee2016}. The proximity to the AF order and spin-orbital coupling give further complexity on the energy and momentum distribution of spin resonance  \cite{chi,qiu,inosov2010,mook,zhang2011,ishikado,jzhao2013,chlee,zhang2013,qureshi2012}. Similar to the cuprate superconductors, the resonance energy in iron-based superconductors overall follows a linear scaling with $T_c$ with slightly different prefactor: $E_R \thickapprox 4.9k_BT_c$, suggesting the common features of the magnetism in various materials and their intimate relation to high-$T_c$ superconductivity [Fig. 1(d)] \cite{pdai2015,inosov2016,mywang2010,gyu2009,masurmach2015,pdjohnson}.

The 112-type iron pnictide Ca$_{1-x}$La$_x$FeAs$_2$ discovered in 2013 has a unique noncentrosymmetric lattice structure derivative from HfCuSi$_2$ with additional zigzag arsenic chains between Ca/La layers \cite{katayama,sjiang2016a}. Its magnetic order is very similar to the collinear antiferromagnetism in in BaFe$_2$As$_2$ (122 type) or NaFeAs (111 type) with same wavevector $\textbf{Q}_{AF}$= (1, 0) in the unit cell with two Fe atoms (or $\textbf{Q}$= (0.5, 0.5) in the one Fe unit cell), but the ordered moments are 45$^\circ$ away from the easy axis of the stripe direction along $a_M$ [Fig. 1(a)] \cite{pdai2015,inosov2016,sjiang2016a,txie}. Comparing to other iron-based superconductors, the fermiology of Ca$_{1-x}$La$_{x}$FeAs$_2$ is also composed of two hole pockets at the zone center ($\Gamma$ point) and one oval-like electron pocket at the zone corner ($M$ point) \cite{sjiang2016a,xliu}. However, one additional 3D hole pocket around $\Gamma$ point and one more electron pocket at the Brillouin zone edge ($X$ point) are contributed by $4p$ orbitals from As chains in hybridization with Fe $3d$ orbitals \cite{myli2015,ztliu2015}.  While filamentary superconductivity can be obtained in the pure La doped compounds \cite{xliu,yakita,hota,kawasaki}, further doping Co or Ni can improve the system to bulk superconductivity with $T_c$ up to 34 K \cite{txie,sjiang2016b}. Unlike the nearly isotropic superconductivity in the 122 system \cite{zswang2015}, transport experiments reveal a quasi-2D behavior of superconducting fluctuations in the 112 system \cite{sonora}.

\begin{figure}[t]
\includegraphics[width=0.4\textwidth]{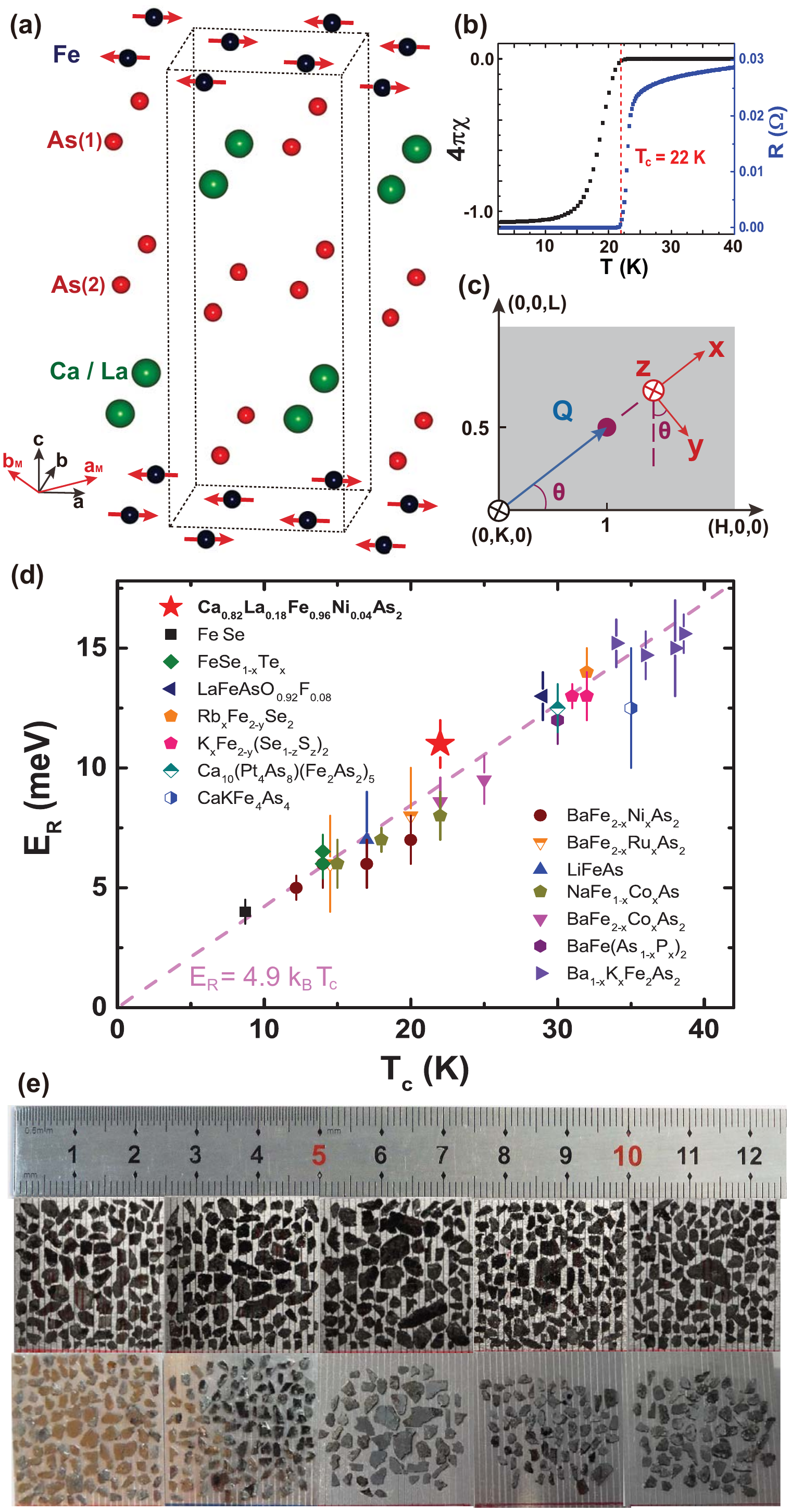}
\caption{
(a) Crystal and magnetic structure of the Ca$_{1-x}$La$_{x}$FeAs$_{2}$ system. (b) Superconducting transitions of our Ca$_{0.82}$La$_{0.18}$Fe$_{0.96}$Ni$_{0.04}$As$_{2}$ samples measured by the resistivity and magnetic susceptibility.
(c) Polarization setup in the reciprocal space for the polarized neutron scattering experiment. (d) Linear scaling between the resonance energy $E_R$ and critical temperature $T_c$ in iron-based superconductors \cite{qswang,qiu,mook,wakimoto,jtpark2011,qswang2,msato2011,kiida,xie2018,pdai2015,chi,mywang2010,jzhao2013,qureshi2012,zhang2013,inosov2010,chlee,zhang2011,chlee2016,pdjohnson}. (e) Photo of the coaligned Ca$_{0.82}$La$_{0.18}$Fe$_{0.96}$Ni$_{0.04}$As$_{2}$ crystals.
}
\end{figure}

\begin{figure}[t]
\includegraphics[width=0.42\textwidth]{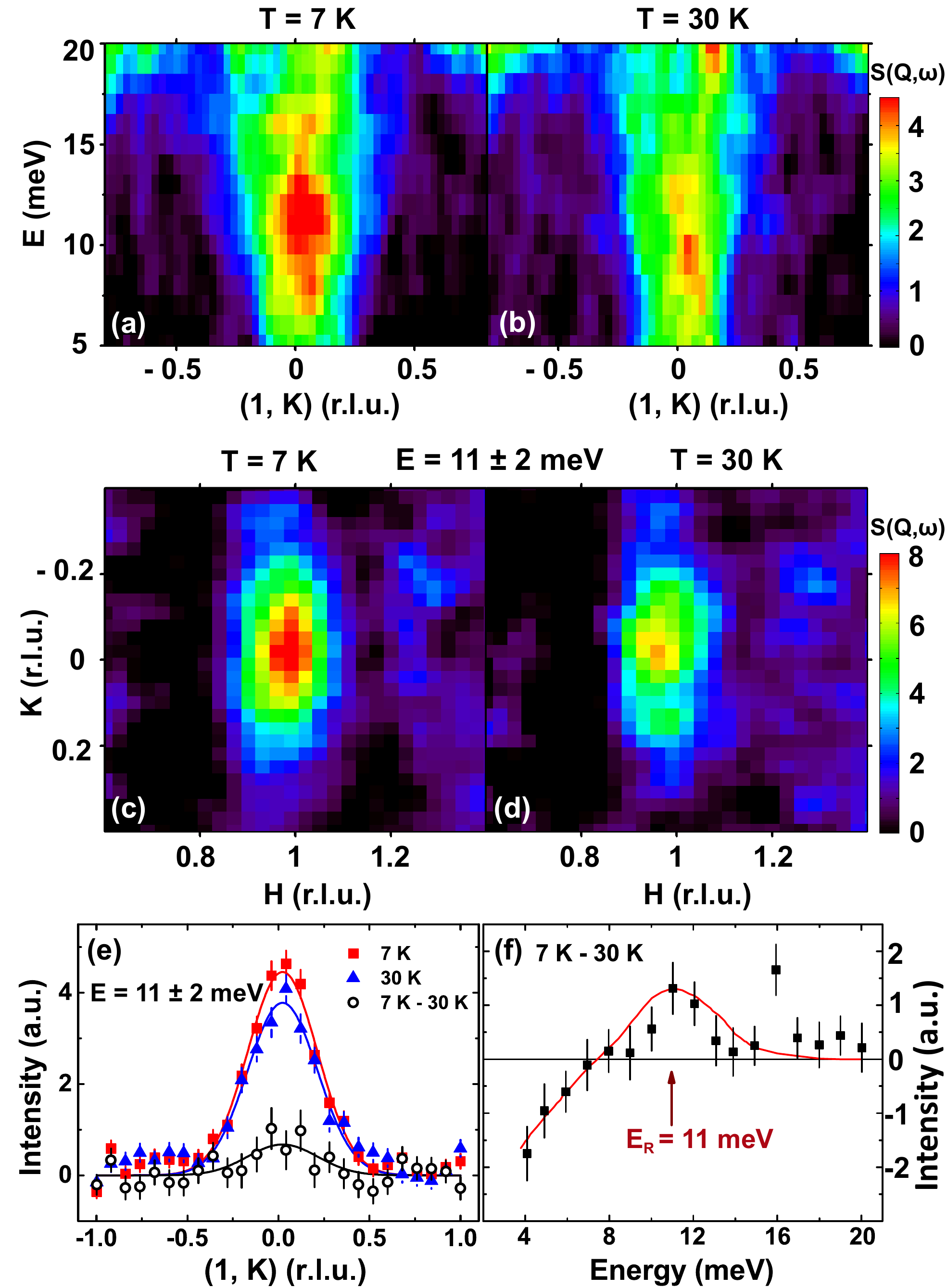}
\caption{(a), (b) Energy dependence of the two-dimensional slices along the $\textbf{Q} = (1, K)$ direction with $E_i = 40$ meV at $T=7$ and 30 K.
(c), (d) Constant-energy slices of the magnetic excitations with $E = 11 \pm 2$ meV at $T=7$ and 30 K.
(e) One-dimensional cuts of the magnetic excitations at $E = 11 \pm 2$ meV along the $\textbf{Q} = (1, K)$ direction. The red and blue lines are Gaussian fits to the raw data, and the black line is their difference.
(f) Spin resonance mode revealed by the difference of the spin excitations between $T=7$ and 30 K. The solid lines are guides to the eyes.
 }
 \end{figure}
Here, in this letter, we report a systematic inelastic neutron scattering study on the 112-compound Ca$_{0.82}$La$_{0.18}$Fe$_{0.96}$Ni$_{0.04}$As$_{2}$ with bulk superconductivity below $T_c=22$ K [Fig. 1(b)].  We have found a 2D spin resonance in reciprocal space with resonance energy $E_R=11$ meV in scaling with other iron-based superconductors [Fig.1 (d)]\cite{christianson,pdai2015,eschrig,lipscombe,slli2010,gyu2009,inosov2016,lumsden,chi,qiu,inosov2010,mook,zhang2011,ishikado,jzhao2013,chlee,zhang2013,qureshi2012,wakimoto,jtpark2011,qswang,qswang2,kiida,xie2018,mwma,msato2011}. Polarized neutron analysis reveals that the resonance is isotropic in spin space due to weak spin-orbit coupling. After comparing with the calculated band structure and Fermi surfaces, we conclude that the spin resonance mode in the 112 system is intrinsically from spin singlet-triplet excitations, and it does not directly respond to the 3D Fermi surface induced by the As $4p$ orbitals.

We prepared high quality single crystals of Ca$_{0.82}$La$_{0.18}$Fe$_{0.96}$Ni$_{0.04}$As$_{2}$ using the self-flux method \cite{txie} and coaligned 2.3 g single crystals [about 1500 pieces, see Fig. 1(e)] with a small mosaic about $4^\circ$ in the $ab$ plane and $3^\circ$ along the $c$ axis  \cite{supplementary} . The transport characterization suggests the superconducting transition is very sharp with $T_c=22$ K and about a 100\% shielding volume at low temperature [Fig. 1(b)]. Time-of-flight neutron scattering experiments were carried out at the HRC spectrometer (BL-12) at J-PARC, Tokai, Japan, with an incident energy $E_i= 40 $ meV and $k_i$ parallel to the $c$ axis.  Unpolarized neutron scattering experiments were performed at the thermal neutron triple-axis spectrometer TAIPAN at Australian Centre for Neutron Scattering, ANSTO, Australia, with fixed final energy $E_f=$ 14.7 meV. The scattering plane $(H, 0, 0) \times (0, 0, L)$ is defined by using a pseudo-orthorhombic magnetic unit cell with $a_M\approx b_M\approx 5.52$ \AA, $c_M=10.27$ \AA, and the vector \textbf{Q} in reciprocal space is defined as $\textbf{Q}=H\textbf{a} ^*+K\textbf{b} ^*+L\textbf{c} ^*$, where $H$, $K$, and $L$ are Miller indices and $\textbf{a} ^*=\hat{\textbf{a}}2\pi/a_M, \textbf{b} ^*=\hat{\textbf{b}}2\pi/b_M, \textbf{c} ^*=\hat{\textbf{c}}2\pi/c_M$ are reciprocal lattice units. Polarized neutron scattering experiments were carried out using the CryoPAD system at the CEA-CRG IN22 thermal triple-axis spectrometer of the Institut Laue-Langevin, Grenoble, France, with the same scattering plane and final energy as the TAIPAN experiment. We define the neutron polarization directions as $x, y, z$, with $x$ parallel to $\textbf{Q}$, and $y$ and $z$ perpendicular to \textbf{Q} as shown in Fig. 1(c). At a specific momentum and energy transfer, magnetically scattered neutrons can have polarizations antiparallel [neutron spin flip (SF) $\uparrow\downarrow$] to the incident neutrons. The three neutron SF scattering cross sections can be written as $\sigma_{\alpha}^{\rm SF}$, where $\alpha=x,y,z$. Since neutron scattering is only sensitive to those magnetic scattering components perpendicular to the momentum transfer \textbf{Q}, we therefore have $\sigma_x^{\textrm{SF}}=cM_y+cM_z+B$, $\sigma_y^{\textrm{SF}}=cM_z+B$,  $\sigma_z^{\textrm{SF}}=cM_y+B$, where $M_y$ and $M_z$ are the magnetic fluctuation moments along the $y$ and $z$ directions, $B$ is the constant background from the instrument, and $c=(R-1)/(R+1)$ with spin flipping ratio $R \approx 15$ in our experiment \cite{mliu2012,hqluo2013,dhu2017,cwang2013,qureshi2012b,ysong2017,ysong2013}. If the spin excitations are completely isotropic in spin space, then $M_y$ must be identical to $M_z$, resulting in $\sigma_y^{\textrm{SF}}=\sigma_z^{\textrm{SF}}=(\sigma_x^{\textrm{SF}}+B)/2$.

\begin{figure}[t]
\includegraphics[width=0.45\textwidth]{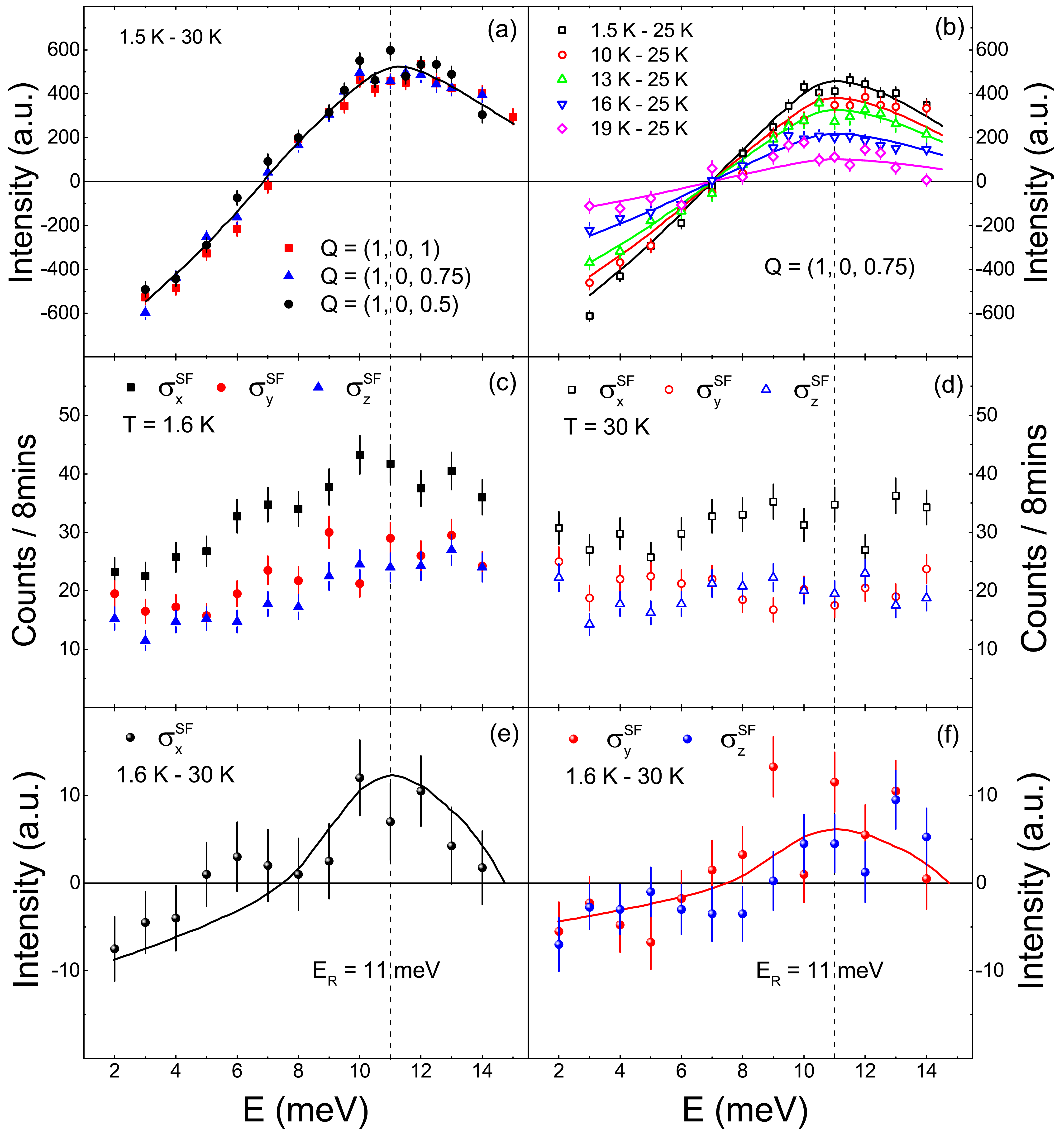}
\caption{ (a) Energy dependence of the spin resonance at $\textbf{Q}=(1, 0, L)$ with $L=$ 0.5, 0.75, and 1 for unpolarized neutron scattering.
(b) Temperature dependence of the spin resonance at $\textbf{Q}=(1, 0, 0.75)$ from 1.5 to 19 K.
(c)-(f) Energy scans at $\textbf{Q}=(1, 0, 0.5)$ for SF scattering below and above $T_c$ and their difference for each neutron polarization direction, marked as $\sigma_{x,y,z}^{\textrm{SF}}$. The solid lines in (a), (b), (d), and (e) are guides to the eyes, and in (f) the solid line is the half of (e).
 }
\end{figure}

\begin{figure}[t]
\includegraphics[width=0.4\textwidth]{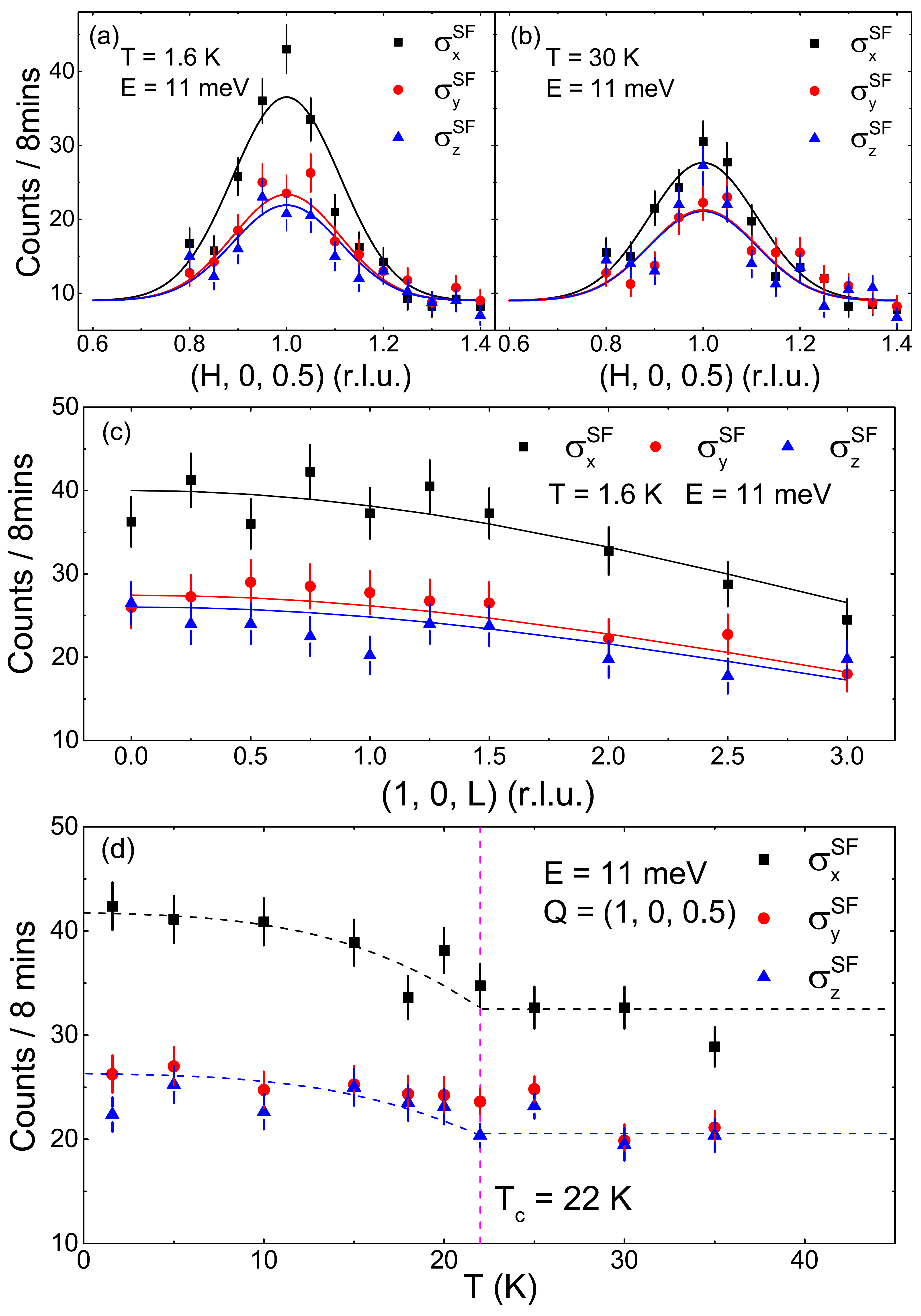}
\caption{
(a), (b) Constant energy scans at $E=$11 meV along $\textbf{Q}=(H, 0, 0.5)$ in the
neutron SF channel $\sigma_{x,y,z}^{\textrm{SF}}$ below and above $T_c$, respectively.
The solid lines are Gaussian fits to the raw data.
(c) Constant energy scans along $\textbf{Q}=(1, 0, L)$  at $E=$11 meV and $T=1.6$ K. The solid lines represent the square of the form factor normalized by the initial intensity for each channel.
(d) Temperature dependence of the neutron SF scattering cross section $\sigma_{x,y,z}^{\textrm{SF}}$
at $E=$11 meV and $\textbf{Q}=(1, 0, 0.5)$. The dashed lines are guides to the eyes.
 }
\end{figure}
Figure 2 shows the energy and momentum dependence of low-energy spin excitations at 7 K (below $T_c$) and 30 K (above $T_c$) measured at HRC. A significant enhancement of the spin excitations from 8 to 15 meV is found at the superconducting state ($T=7$ K), with an ellipse shape elongated along the $K$ direction but no clear energy dispersion [Figs. 2(a)-2(d)]. By comparing the spin excitation intensity below and above $T_c$, a resonance mode is identified to be around 11 meV at the zone center $\textbf{Q}=(1, 0)$ (Figs. 2(e) and 2(f)). Given $T_c=22$ K, we have $E_R = 5.8k_BT_c$, which is close to the prefactor 4.9 as shown in Fig. 1(d) \cite{pdjohnson}. The momentum transfer and temperature dependence of the spin resonance is measured at TAIPAN, as shown in Figs. 3(a) and 3(b). For clarity, we only present the intensity differences between the superconducting and normal states. The spin excitations remain almost the same for different $L=$ 0.5, 0.75, and 1 from 2 to 15 meV within the first Brillouin zone, suggesting the 2D nature of the resonance mode. Moreover, the resonance energy corresponding to the maximum intensity is nearly temperature independent; thus, it does not exactly follow the temperature dependence of the BCS-like superconducting gaps \cite{inosov2010,zhang2011,harriger2012,hqluo2013b,mazin2009,maier2008,maier2009,korshunov2008,zhang2013b}.

To further check the anisotropy of the spin resonance in energy, $\textbf{Q}$, and the temperature dependence, we further carried out polarized neutron scattering at IN22. The raw data of the magnetic scattering in SF channels at $T=$ 1.6 K (below $T_c$) and 30 K (above $T_c$) is shown in Figs. 3(c) and 3(d), respectively. There is no clear difference between $\sigma_y^{\textrm{SF}}$ and $\sigma_z^{\textrm{SF}}$ at both temperatures within error bars, suggesting isotropic spin excitations in both superconducting and normal states \cite{mliu2012,hqluo2013,dhu2017}. The difference of $\sigma_{\alpha}^{\rm SF}$ ($\alpha=x,y,z$) at $\textbf{Q}=(1, 0, 0.5)$ below and above $T_c$ confirms the 11 meV resonance mode [Figs. 3(e) and 3(f)]. Since the temperature difference plots in Figs. 3(e) and 3(f) should contain no background, we expect $\sigma_{x}^{\rm SF}=\sigma_{y}^{\rm SF}+\sigma_{z}^{\rm SF}=2\sigma_{y}^{\rm SF}$ \cite{hqluo2013}. The red line in Fig. 3(f) indicates the half intensity of $\sigma_{x}^{\rm SF}$ (black line) in Fig. 3(e), and it is indeed statistically identical to $\sigma_{y}^{\rm SF}$ or $\sigma_{z}^{\rm SF}$.

\begin{figure}[t]
\includegraphics[width=0.45\textwidth]{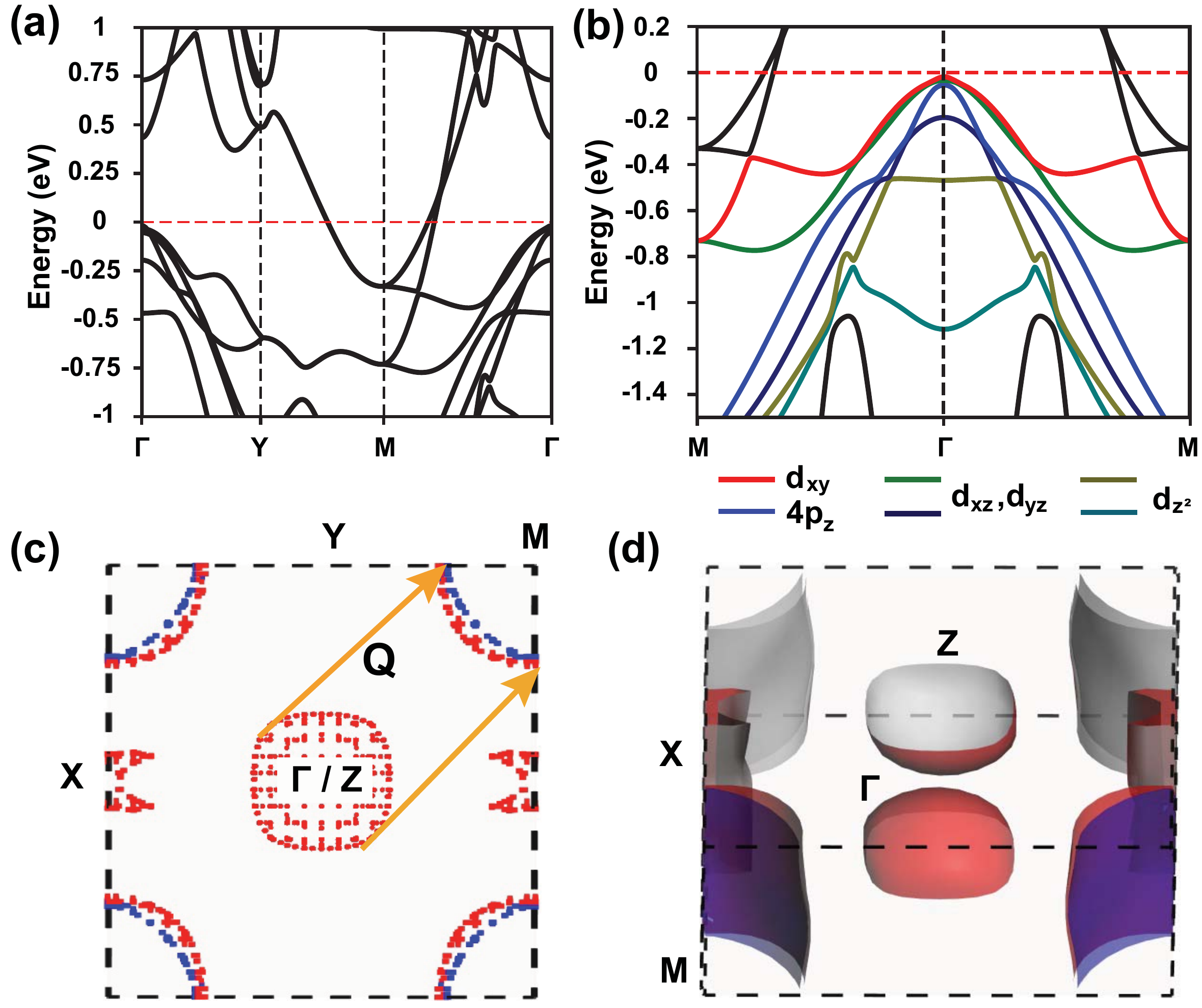}
\caption{
(a), (b) Band structure and orbital occupation of Ca$_{0.82}$La$_{0.18}$Fe$_{0.96}$Ni$_{0.04}$As$_{2}$ from the GGA calculation.
(c), (d) Fermi surfaces in 2D projection and 3D reciprocal space from the GGA calculation.
 }
\end{figure}

Figures 4(a) and 4(b) summarize constant energy scans along the $(H, 0, 0.5)$ direction at the resonance energy $E_R=11$ meV with different neutron polarizations. All three channels display well-defined Gaussian peaks both at $T=1.6$  K and 30 K. After subtracting the flat background, the spin excitations are nearly isotropic by satisfying $\sigma_{x}^{\rm SF}=2\sigma_{y}^{\rm SF}=2\sigma_{z}^{\rm SF}$, except for a negligible difference between $\sigma_{y}^{\rm SF}$ and $\sigma_{z}^{\rm SF}$ at the zone center with $T=1.6$ K. Since the magnetic fluctuating moments can be expressed by $M_y=M_a\sin^2\theta+M_c\cos^2\theta$ and $M_z=M_b$ [$\theta$ is the angle between $\textbf{Q}$ and the $(H, 0, 0)$ directions] \cite{lipscombe,hqluo2013,dhu2017,ysong2013}, and as we actually have $M_a=M_b$ due to the diagonal orientation of ordered moments \cite{txie,sjiang2016b}, the difference between $M_y$ and $M_z$ ($\sigma_{y}^{\rm SF}$ and $\sigma_{z}^{\rm SF}$) must come from the anisotropic excitations between $M_a$ and $M_c$ and decrease with increasing $L$. It turns out to be not the case in our experiment, because all three channels of the spin resonance do not have either any clear modulation or an abnormal dependence besides the intensity following the square of the form factor $\mid f_M(\textbf{Q})\mid ^2$ ranging from $L=0$ to $L=3$ [Fig. 4(c)]. The small and persistent difference between $\sigma_{y}^{\rm SF}$ and $\sigma_{z}^{\rm SF}$ in Fig. 4(c) is attributed to the slight change of the flipping ratio in these two channels instead of the spin anisotropy. Finally, a temperature dependence of the spin resonance at all three neutron polarizations is shown in Fig. 4(d). The intensity of $\sigma_{x}^{\rm SF}$ likes an order parameter and responds to $T_c$, and the spin excitations are isotropic except for the perturbation from critical scattering around $T_c$ \cite{ysong2017,ysong2013}.

To figure out the band structure and fermiology in this particular compound, we performed generalized-gradient-approximation (GGA) calculations, as shown in Fig. 5. Similar to the Ni free system Ca$_{1-x}$La$_x$FeAs$_2$, the As $4p$ orbitals are involved besides the Fe $3d$ orbitals (Figs. 5(a) and 5(b)), resulting in a 3D hole pocket around the $\Gamma$ point and a small 2D electron pocket around the $X$ point (Figs. 5(c) and 5(d)). The other 2D hole sheet outside the zone center, as shown in the dynamical mean field theory (DMFT) calculation of Ca$_{1-x}$La$_x$FeAs$_2$ \cite{sjiang2016a,xliu,myli2015,ztliu2015}, is not found in our GGA calculation due to the Ni doping effect. Since we only find a 2D spin resonance around $\textbf{Q}=(1, 0)$ (Fig. 2) corresponding to the wave vector $\textbf{Q}$ from $\Gamma$/$Z$ to $M$ [Fig. 5(c)], the extra electron pocket at the $X$ point thus has no contribution to the spin excitations or superconductivity. The $s\pm$ pairing only happens on the 3D hole pocket around the $\Gamma$ point and 2D electron pockets at the $M$ point.

The complex multiorbital nature in iron-based superconductors may lead to a dispersion and spin anisotropy of the spin resonance \cite{hqluo2013,steffens2013,mwang2016,chlee,dhu2017,zhang2013b,zhang2014,wwang2017,babkevich,mwma2017,ysong2016}. In a recent study on CaKFe$_4$As$_4$ (1144 compound), the spin resonance is found to be composed with nondispersive triple modes with energies scaling with the total superconducting gaps summed on the nesting hole and electron pockets. In contrast to its 2D fermiology, the resonance intensity has highly 3D modulation with both odd and even $L$ symmetries due to the nondegenerate spin excitations from the Fe-As bilayer \cite{xie2018}. For our case in the 112 system, although the broad energy distribution of the spin resonance may be involved with multimodes due to the double electron sheets, the resonance energy and intensity is completely $L$ indepdent, suggesting a 2D nature of the resonant mode despite the 3D hole pocket.  Therefore, the spin resonance actually does not respond to the $k_z$ dependence of the fermiology in the iron-based superconductor. More uniquely for the 112 system, the spin-orbit coupling is rather weak due to the decoupled structural transition and magnetic transition in the \textquotedblleft parent\textquotedblright compound \cite{sjiang2016a,txie}; the Ni doping further weakens the orbital ordering as well as the nematicity \cite{yhgu,ckang2017}. This fact enables us to reveal that the spin resonance intrinsically is an isotropic spin-1 collective mode of spin excitations. Moreover, the presence of additional zigzag As chains probably blocks the correlations for FeAs interlayers; then the Ni doping can eliminate the 3D AF order very easily \cite{txie}, giving rise to a 2D superconductivity and spin resonance in the doped compounds \cite{sonora,xing2016}. Finally, the nearly temperature independent resonance energy is also consistent with the $s\pm$ pairing mechanism \cite{zhang2013b}.

In summary, inelastic neutron scattering experiments have been carried out on the 112-type iron pnictide superconductor Ca$_{0.82}$La$_{0.18}$Fe$_{0.96}$Ni$_{0.04}$As$_{2}$. A 2D spin resonance mode around $E=11$ meV is found. The nearly temperature independent resonance energy scales with $T_c$ along with other iron-based superconductors. The isotropic intensity of the resonance in spin space suggests that it is a spin-1 excited mode of Cooper pairs in the case of weak spin-orbital coupling. Comparing with our recent study on CaKFe$_4$As$_4$ \cite{xie2018}, we argue that the spin resonance in the iron-based superconductor intrinsically is a spin singlet-triplet exciton and does not respond to the $k_z$ dependence of the fermiology.

This work is supported by the National Natural Science Foundation of China (Grants No. 11374011, No. 11374346, No. 11674406, and No. 11674372), the National Key Research and Development Program of China (Grants No. 2017YFA0302903, No. 2017YFA0303103, No. 2016YFA0300502) and the Strategic Priority Research Program (B) of the Chinese Academy of Sciences (Grants No. XDB07020300, No. XDPB01). H. L. is supported by the Youth Innovation Promotion Association of CAS. A.G. is supported by HBNI, and H.G. gratefully acknowledges P. A. Naik, A. Banerjee for support. The neutron-scattering experiment at HRC was performed with the J-PARC (2016B0068).

\clearpage
\begin{center}
{\bf Supplementary Materials}
\end{center}

We prepared high quality single crystals of Ca$_{0.82}$La$_{0.18}$Fe$_{0.96}$Ni$_{0.04}$As$_{2}$ using self-flux method, the detailed procedure of the crystal growth is shown in our previous report \cite{txie}. The size scale of the crystals is about 1 to 6 mm. We co-aligned 2.3 g single crystals (about 1500 pieces) by a X-ray Laue camera (\emph{Photonic Sciences}) in backscattering mode with incident beam along $c$-axis. A hydrogen-free glue named \emph{CYTOP} was painted on the back side of the crystals by sticking on thin aluminium plates and baked for 1-2 hours at 100 $^\circ$C. The scattering plane was set up as $(H, 0, 0) \times (0, 0, L)$ with several aluminium plates with samples (Fig. S1). The neutron diffraction measurements on the whole sample array reveal a small mosaic of these co-aligned crystals. The rocking curves can be well fitted by Gaussian functions with full-width-at-half-maximum (FWHM) less than $3.7^\circ$ for $ab-$plane($Q = (2, 0, 0)$) and $2.8^\circ$ along $c-$axis($Q = (0, 0, 4)$) (Fig. S2).

Figure S3(a) shows the normalized resistivity ($\rho/\rho_{300 K}$) of our Ca$_{0.82}$La$_{0.18}$Fe$_{0.96}$Ni$_{0.04}$As$_{2}$ crystals. The sharp superconducting transitions, uniform $T_c$ and nearly identical normal state behaviors among 8 randomly selected samples indicate high quality and homogeneity of our samples. No kinks of $\rho$ related to the structural or magnetic transition can be found. Figure S3(b) shows the temperature dependence of the DC magnetic susceptibility for 2 typical crystals picked up from the resistivity measurements. The superconducting transitions are very sharp, and both of the samples have full Meissner shielding volume ($4\pi\chi\approx -1$) at based temperature ($T = 2$ K), suggesting bulk superconductivity in our crystals.

\newpage
\begin{figure*}[t]
\renewcommand\thefigure{S1}
\includegraphics[width=0.25\textwidth]{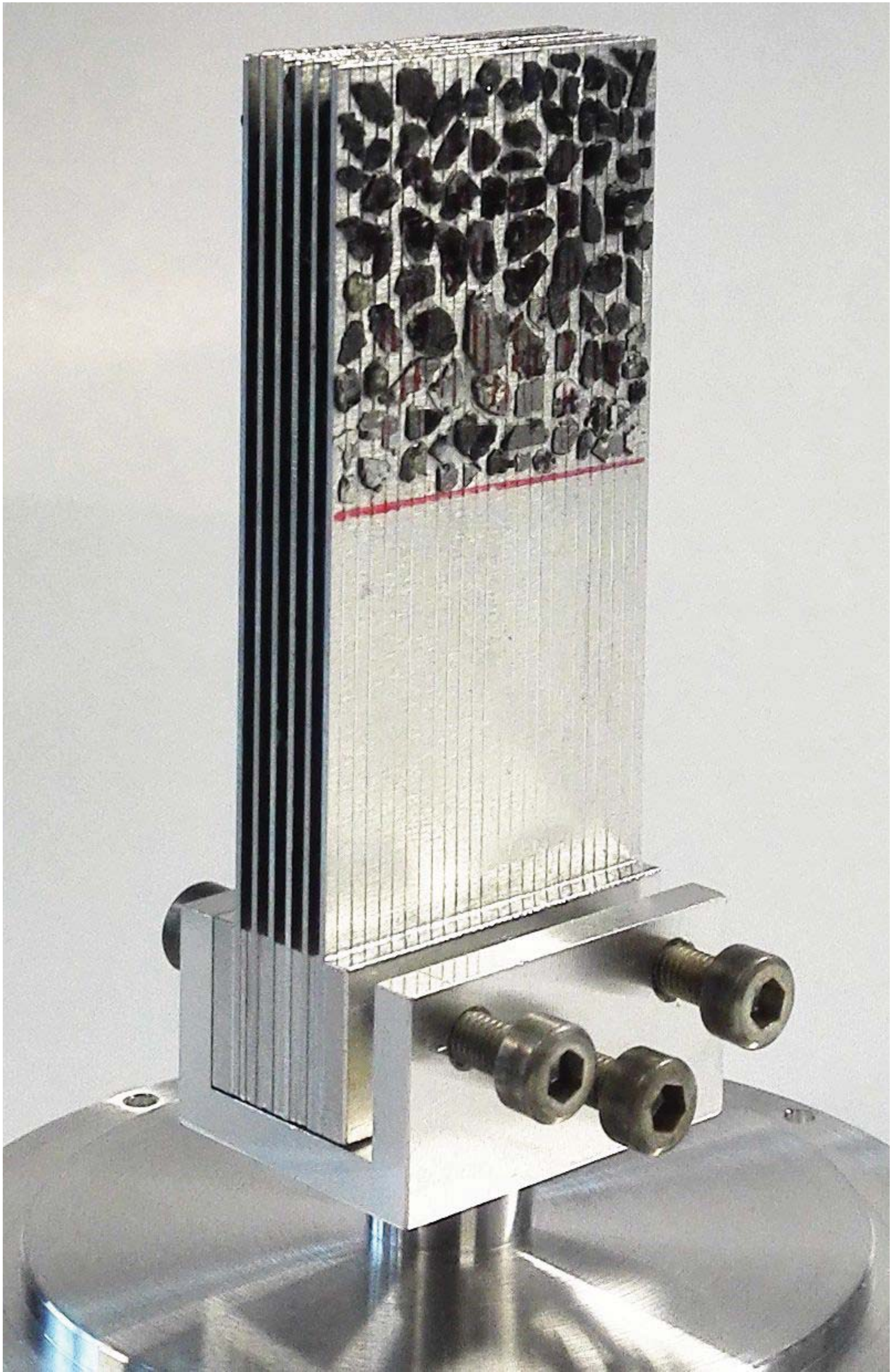}
\caption{
Photos of the assembled Ca$_{0.82}$La$_{0.18}$Fe$_{0.96}$Ni$_{0.04}$As$_{2}$ crystals on for neutron scattering experiments.
}
\end{figure*}

\begin{figure*}[t]
\renewcommand\thefigure{S2}
\includegraphics[width=0.8\textwidth]{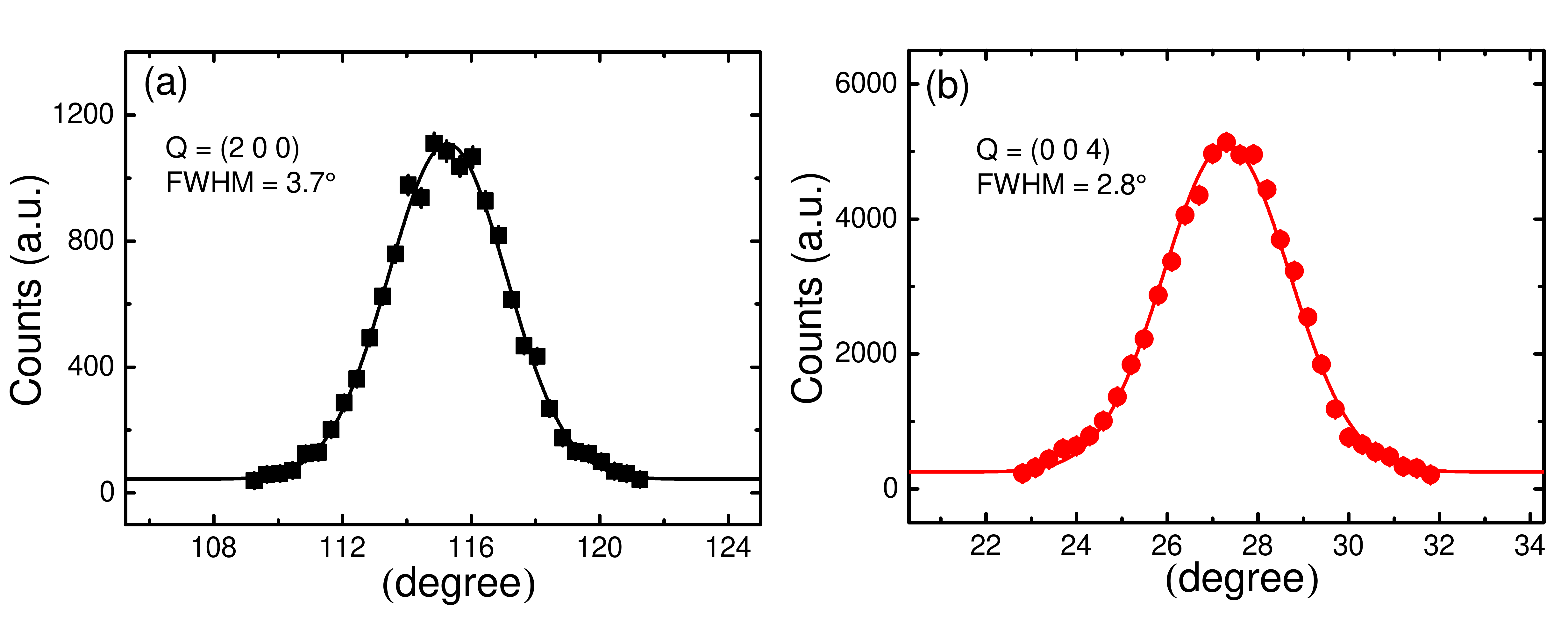}
\caption{ Rocking curves of co-aligned crystals measured by neutron diffraction experiments. The solid lines are Gaussian fits of the data.
 }
 \end{figure*}

\begin{figure*}[t]
\renewcommand\thefigure{S3}
\includegraphics[width=0.8\textwidth]{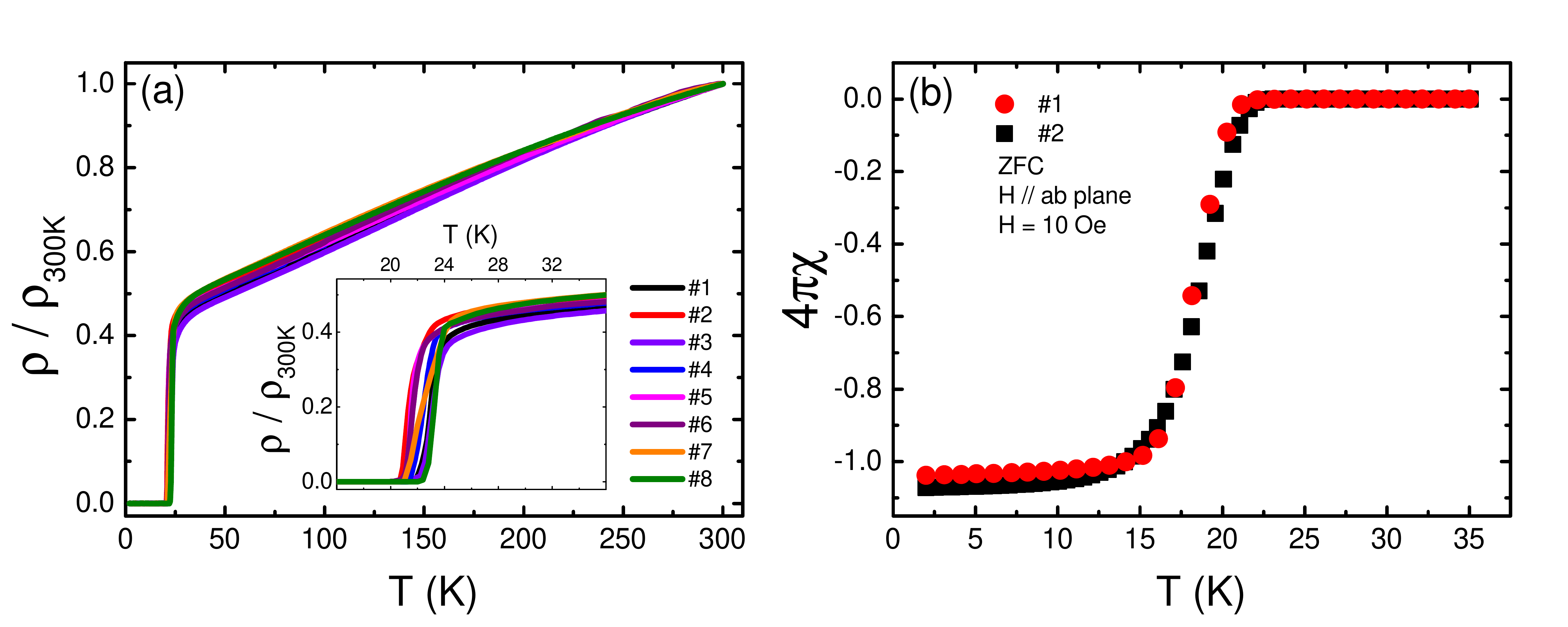}
\caption{Transport characterizations of our single crystals: (a) Temperature dependence of the resistivity, all the data is normalized by the resistivity at 300 K; (b) Temperature dependence of DC magnetic susceptibility.
 }
\end{figure*}


\begin{thebibliography}{}
\bibitem{rossat-mignod} J. Rossat-Mignod, L.-P. Regnault, C. Vettier, P. Bourges, P. Burlet, J. Bossy, J. Y. Henry, and G. Lapertot, Physica (Amsterdam) {\bf 185C}-{\bf189C}, 86 (1991).
\bibitem{nksato} N. K. Sato, N. Aso, K. Miyake, R. Shiina, P. Thalmeier, G. Varelogiannis, C. Geibel, F. Steglich, P. Fulde, and T. Komatsubara, Nature (London) {\bf 410}, 340 (2001).
\bibitem{wilson} S. D. Wilson, P. Dai, S. Li, S. Chi, H. J. Kang, and J. W. Lynn, Nature (London) {\bf 442}, 59 (2006).
\bibitem{christianson} A. D. Christianson, E. A. Goremychkin, R. Osborn, S. Rosenkranz, M. D. Lumsden, C. D. Malliakas, I. S. Todorov, H. Claus, D. Y. Chung, M. G. Kanatzidis, R. I. Bewley, and  T. Guidi, Nature (London) {\bf 456}, 930 (2008).
\bibitem{pdai2015} P. Dai, Rev. Mod. Phys. {\bf 87}, 855 (2015).
\bibitem{eschrig} M. Eschrig, Adv. Phys. {\bf 55}, 47 (2006).
\bibitem{lipscombe} O. J. Lipscombe, L. W. Harriger, P. G. Freeman, M. Enderle, C. Zhang, M. Wang, T. Egami, J. Hu, T. Xiang, M. R. Norman, and P. Dai, Phys. Rev. B {\bf 82}, 064515 (2010).
\bibitem{slli2010} S. Li, X. Lu, M. Wang, H. Q. Luo, M. Wang, C. Zhang, E. Faulhaber, L.-P. Regnault, D. Singh, and P. Dai, Phys. Rev. B {\bf 84}, 024518 (2011).
\bibitem{gyu2009} G. Yu, Y. Li, E. M. Motoyama, and M. Greven, Nat. Phys. {\bf 5}, 873 (2009).
\bibitem{inosov2016} D. S. Inosov, C. R. Phys. {\bf 17}, 60 (2016).
\bibitem{lumsden} M. D. Lumsden, A. D. Christianson, D. Parshall, M. B. Stone, S. E. Nagler, G. J. MacDougall, H. A. Mook, K. Lokshin, T. Egami, D. L. Abernathy, E. A. Goremychkin, R. Osborn, M. A. McGuire, A. S. Sefat, R. Jin, B. C. Sales, and D. Mandrus, Phys. Rev. Lett. {\bf 102}, 107005 (2009).
\bibitem{chi} S. Chi, A. Schneidewind, J. Zhao, L. W. Harriger, L. Li, Y. Luo, G. Cao, Z. Xu, M. Loewenhaupt, J. Hu, and P. Dai, Phys. Rev. Lett. {\bf 102}, 107006 (2009).
\bibitem{qiu} Y. Qiu, W. Bao, Y. Zhao, C. Broholm, V. Stanev, Z. Tesanovic, Y. C. Gasparovic, S. Chang, J. Hu, B. Qian, M. Fang, and Z. Mao, Phys. Rev. Lett. {\bf 103}, 067008 (2009).
\bibitem{mook} H. A. Mook, M. D. Lumsden, A. D. Christianson, S. E. Nagler, B. C. Sales, R. Jin, M. A. McGuire, A. S. Sefat, D. Mandrus, T. Egami, and C. dela. Cruz, Phys. Rev. Lett. {\bf 104}, 187002 (2010).
\bibitem{inosov2010} D. S. Inosov, J. T. Park, P. Bourges, D. L. Sun, Y. Sidis, A. Schneidewind, K. Hradil, D. Haug, C. T. Lin, B. Keimer, and V. Hinkov, Nat.Phys. {\bf 6}, 178 (2010).
\bibitem{zhang2011} C. Zhang, M. Wang, H. Luo, M. Wang, M. Liu, J. Zhao, D. L. Abernathy, T. A. Maier, K. Marty, M. D. Lumsden, S. Chi, S. Chang, J. A. Rodriguez-Rivera, J. W. Lynn, T. Xiang, J. Hu, and P. Dai, Sci. Rep. {\bf 1}, 115 (2011).
\bibitem{ishikado} M. Ishikado, Y. Nagai, K. Kodama, R. Kajimoto, M. Nakamura, Y. Inamura, S. Wakimoto, H. Nakamura, M. Machida, K. Suzuki, H. Usui, K. Kuroki, A. Iyo, H. Eisaki, M. Arai, and S. I. Shamoto, Phys. Rev. B {\bf 84}, 144517 (2011).
\bibitem{jzhao2013} J. Zhao, C. R. Rotundu, K. Marty, M. Matsuda, Y. Zhao, C. Setty, E. Bourret-Courchesne, J. Hu, and R. J. Birgeneau, Phys. Rev. Lett. {\bf 110}, 147003 (2013).
\bibitem{chlee} C. H. Lee, P. Steffens, N. Qureshi, M. Nakajima, K. Kihou, A. Iyo, H. Eisaki, and M. Braden, Phys. Rev. Lett. {\bf 111}, 167002 (2013).
\bibitem{zhang2013}	C. Zhang, R. Yu, Y. Su, Y. Song, M. Wang, G. Tan, T. Egami, J. A. Fernandez-Baca, E. Faulhaber, Q. Si, and P. Dai, Phys. Rev. Lett. {\bf 111}, 207002 (2013).
\bibitem{qureshi2012} N. Qureshi, P. Steffens, Y. Drees, A. C. Komarek, D. Lamago, Y. Sidis, L. Harnagea, H.-J. Grafe, S. Wurmehl, B. B\"{u}chner, and M. Braden, Phys. Rev. Lett. {\bf 108}, 117001 (2012).
\bibitem{wakimoto} S. Wakimoto, K. Kodama, M. Ishikado, M. Matsuda, R. Kajimoto, M. Arai, K. Kakurai, F. Esaka, A. Iyo, H. Kito, H. Eisaki, and S. Shamoto, J. Phys. Soc. Jpn, {\bf 79}, 074715 (2010).
\bibitem{jtpark2011} J. T. Park, G. Friemel, Yuan Li, J.-H. Kim, V. Tsurkan, J. Deisenhofer, H.-A. Krug von Nidda, A. Loidl, A. Ivanov, B. Keimer, and D. S. Inosov, Phys. Rev. Lett. {\bf 107}, 177005 (2011).
\bibitem{qswang} Q. Wang, Y. Shen, B. Pan, Y. Hao, M. Ma, F. Zhou, P. Steffens, K. Schmalzl, T. R. Forrest, M. Abdel-Hafiez, X. Chen, D. A. Chareev, A. N. Vasiliev, P. Bourges, Y. Sidis, H. Cao, and J. Zhao, Nat. Mater. {\bf 15}, 159 (2016).
\bibitem{kiida} K. Iida, M. Ishikado, Y. Nagai, H. Yoshida, A. D. Christianson, N. Murai, K. Kawashima, Y. Yoshida, H. Eisaki, and A. Iyo, J. Phys. Soc. Jpn. {\bf 86}, 093703 (2017).
\bibitem{xie2018} T. Xie, Y. Wei, D. Gong, T. Fennell, U. Stuhr, R. Kajimoto, K. Ikeuchi, S. Li, J. Hu, and H. Luo, arXiv:1802.01901.
\bibitem{mwma} M. Ma, L. Wang, P. Bourges, Y. Sidis, S. Danilkin, and Y. Li, Phys. Rev. B {\bf 95}, 100504 (2017).
\bibitem{msato2011} M. Sato, T. Kawamata, Y. Kobayashi, Y. Yasui, T. Iida, K. Suzuki, M. Itoh, T. Moyoshi, K. Motoya, R. Kajimoto, M. Nakamura, Y. Inamura, and M. Arai, J. Phys. Soc. Jpn. {\bf 80}, 093709 (2011).
\bibitem{qswang2} Q. Wang, J. T. Park, Y. Feng, Y. Shen, Y. Hao, B. Pan, J. W. Lynn, A. Ivanov, S. Chi, M. Matsuda, H. Cao, R. J. Birgeneau, D. V. Efremov, and J. Zhao, Phys. Rev. Lett. {\bf 116}, 197004 (2016).
\bibitem{masurmach2015} M. A. Surmach, F. Br\"{u}ckner, S. Kamusella, R. Sarkar, P. Y. Portnichenko, J. T. Park, G. Ghambashidze, H. Luetkens, P. K. Biswas, W. J. Choi, Y. I. Seo, Y. S. Kwon, H.-H. Klauss, and D. S. Inosov, Phys. Rev. B {\bf 91}, 104515 (2015).
\bibitem{kikeuchi2014} K. Ikeuchi, M. Sato, R. Kajimoto, Y. Kobayashi, K. Suzuki, M. Itoh, P. Bourges, A. D. Christianson, H. Nakamura, and M. Machida, JPS Conf. Proc. {\bf 3}, 015043 (2014).
\bibitem{chlee2016} C. H. Lee, K. Kihou, J. T. Park, K. Horigane, K. Fujita, F. Wa{\ss}er, N. Qureshi, Y. Sidis, J. Akimitsu, and M. Braden, Sci. Rep. {\bf 6}, 23424 (2016).
\bibitem{mazin2008} I. I. Mazin, D. J. Singh, M. D. Johannes, and M. H. Du, Phys. Rev. Lett. {\bf 101}, 057003 (2008).
\bibitem{kuroki2008} K. Kuroki, S. Onari, R. Arita, H. Usui, Y. Tanaka, H. Kontani, and H. Aoki, Phys. Rev. Lett. {\bf 101}, 087004 (2008).
\bibitem{ding2008} H. Ding, P. Richard, K. Nakayama, K. Sugawara, T. Arakane, Y. Sekiba, A. Takayama, S. Souma, T. Sato, and T. Takahashi, Europhys. Lett. {\bf 83}, 47001 (2008).
\bibitem{mywang2010} M. Wang, H. Luo, J. Zhao, C. Zhang, M. Wang, K. Marty, S. Chi, J. W. Lynn, A. Schneidewind, S. Li, and P. Dai, Phys. Rev. B {\bf 81}, 174524 (2010).
\bibitem{pdjohnson} P. D. Johnson, G. Xu, and W. -G. Yin {\it Iron-Based Superconductivity}, (Springer, New York, 2015), pp 165-169.
\bibitem{katayama} N. Katayama, K. Kudo, S. Onari1, T. Mizukami, K. Sugawara, Y. Sugiyama, Y. Kitahama, K. Iba, K. Fujimura, N. Nishimoto, M. Nohara, and H. Sawa, J. Phys. Soc. Jpn. {\bf 82}, 123702 (2013).
\bibitem{sjiang2016a} S. Jiang, C. Liu, H. Cao, T. Birol, J. M. Allred, W. Tian, L. Liu, K. Cho, M. J. Krogstad, J. Ma, K. M. Taddei, M. A. Tanatar, M. Hoesch, R. Prozorov, S. Rosenkranz, Y. J. Uemura, G. Kotliar, and N. Ni, Phys. Rev. B. {\bf 93}, 054522 (2016).
\bibitem{txie} T. Xie, D. Gong, W. Zhang, Y. Gu, Z. Huesges, D. Chen, Y. Liu, L. Hao, S. Meng, Z. Lu, S. Li, and H. Luo, Supercond. Sci. Technol. {\bf 30}, 095002 (2017).
\bibitem{xliu} X. Liu, D. Liu, L. Zhao, Q. Guo, Q. Mu, D. Chen, B. Shen, H. Yi, J. Huang, J. He, Y. Peng, Y. Liu, S. He, G. Liu, X. Dong, J. Zhang, C. Chen, Z. Xu, Z. Ren, and X. Zhou, Chin. Phys. Lett. {\bf 30}, 127402 (2013).
\bibitem{myli2015} M. Y. Li, Z. T. Liu, W. Zhou, H. F. Yang, D. W. Shen, W. Li, J. Jiang, X. H. Niu, B. P. Xie, Y. Sun, C. C. Fan, Q. Yao, J. S. Liu, Z. X. Shi, and X. M. Xie, Phys. Rev. B {\bf 91}, 045112 (2015).
\bibitem{ztliu2015} Z. Liu, X. Xing, M. Li, W. Zhou, Y. Sun, C. Fan, H. Yang, J. Liu, Q. Yao, W. Li, Z. Shi, D. Shen, and Z. Wang, Appl. Phys. Lett. {\bf 109}, 042602 (2016).
\bibitem{yakita} H. Yakita, H. Ogino, T. Okada, A. Yamamoto, K. Kishio, T. Tohei, Y. Ikuhara, Y. Gotoh, H. Fujihisa, K. Kataoka, H. Eisaki, and J. Shimoyama, J. Am. Chem. Soc.,{\bf 136}, 846 (2014).
\bibitem{hota} H. Ota, K. Kudo, T. Kimura, Y. Kitahama, T. Mizukami, S. Ioka, and M. Nohara, J. Phys. Soc. Jpn. {\bf 86}, 025002 (2017).
\bibitem{kawasaki} S. Kawasaki, T. Mabuchi, S. Maeda, T. Adachi, T. Mizukami, K. Kudo, M. Nohara, and G. Q. Zheng, Phys. Rev. B {\bf 92}, 180508 (2015).
\bibitem{sjiang2016b} S. Jiang, L. Liu, M. Sch\"{u}tt, A. M. Hallas, B. Shen, W. Tian, E. Emmanouilidou, A. Shi, G. M. Luke, Y. J. Uemura, R. M. Fernandes, and N. Ni, Phys. Rev. B. {\bf 93}, 174513 (2016).
\bibitem{zswang2015} Z. Wang, T. Xie, E. Kampert, T. F\"{o}rster, X. Lu, R. Zhang, D. Gong, S. Li, T. Herrmannsd\"{o}rfer, J. Wosnitza, and H. Luo, Phys. Rev. B. {\bf 92}, 174509 (2015).
\bibitem{sonora} D. S\'{o}\~{n}ora, C. Carballeira, J. J. Ponte, T. Xie, H. Luo, S. Li, and J. Mosqueira, Phys. Rev. B {\bf 96}, 014516 (2017).
\bibitem{supplementary} See Supplemental Material, which includes the sample photo and characterizations.
\bibitem{mliu2012} M. Liu, C. Lester, J. Kulda, X. Lu, H. Luo, M. Wang, S. M. Hayden, and P. Dai,  Phys. Rev. B {\bf 85}, 214516 (2012).
\bibitem{hqluo2013} H. Luo, M. Wang, C. Zhang, X. Lu, L.-P. Regnault, R. Zhang, S. Li, J. Hu, and P. Dai, Phys. Rev. Lett. {\bf 111}, 107006 (2013).
\bibitem{dhu2017} D. Hu, W. Zhang, Y. Wei, B. Roessli, M. Skoulatos, L.-P. Regnault, G. Chen, Y. Song, H. Luo, S. Li, and P. Dai, Phys. Rev. B {\bf 96}, 180503(R) (2017).
\bibitem{cwang2013} C. Wang, R. Zhang, F. Wang, H. Luo, L.-P. Regnault, P. Dai, and Y. Li,  Phys. Rev. X {\bf 3}, 041036 (2013).
\bibitem{qureshi2012b} N. Qureshi, P. Steffens, S. Wurmehl, S. Aswartham, B. B\"{u}chner, and M. Braden, Phys. Rev. B {\bf 86}, 060410(R) (2012).
\bibitem{ysong2017} Y. Song, W. Wang, C. Zhang, Y. Gu, X. Lu, G. Tan, Y. Su, F. Bourdarot, A. D. Christianson, S. Li, and P. Dai, Phys. Rev. B {\bf 96}, 184512 (2017).
\bibitem{ysong2013} Y. Song, L.-P. Regnault, C. Zhang, G. Tan, S. V. Carr, S. Chi, A. D. Christianson, T. Xiang, and P. Dai, Phys. Rev. B {\bf 88}, 134512 (2013).
\bibitem{harriger2012} L. W. Harriger, O. J. Lipscombe, C. Zhang, H. Luo, M. Wang, K. Marty, M. D. Lumsden, and P. Dai, Phys. Rev. B {\bf 85}, 054511 (2012).
\bibitem{hqluo2013b} H. Luo, X. Lu, R. Zhang, M. Wang, E. A. Goremychkin, D. T. Adroja, S. Danilkin, G. Deng, Z. Yamani, and P. Dai, Phys. Rev. B. {\bf 88}, 144516 (2013).
\bibitem{mazin2009} I. I. Mazin and J. Schmalian, Physica (Amsterdam) {\bf 469C}, 614 (2009).
\bibitem{maier2008} T. A. Maier and D. J. Scalapino, Phys. Rev. B {\bf 78}, 020514(R) (2008).
\bibitem{maier2009} T. A. Maier, S. Graser, D. J. Scalapino, and P. Hirschfeld, Phys. Rev. B {\bf 79}, 134520 (2009).
\bibitem{korshunov2008} M. M. Korshunov and I. Eremin, Phys. Rev. B  {\bf 78}, 140509(R) (2008).
\bibitem{zhang2013b} C. Zhang, H. F. Li, Y. Song, Y. Su, G. Tan, T. Netherton, C. Redding, S. V. Carr, O. Sobolev, A. Schneidewind, E. Faulhaber, L. W. Harriger, S. Li, X. Lu, D. X. Yao, T. Das, A. V. Balatsky, Th. Br\"{u}ckel, J. W. Lynn, and P. Dai, Phys. Rev. B {\bf 88}, 064504 (2013).
\bibitem{mwang2016} M. Wang, M. Yi, H. L. Sun, P. Valdivia, M. G. Kim, Z. J. Xu, T. Berlijn, A. D. Christianson, Songxue Chi, M. Hashimoto, D. H. Lu, X. D. Li, E. Bourret-Courchesne, Pengcheng Dai, D. H. Lee, T. A. Maier, and R. J. Birgeneau, Phys. Rev. B {\bf 93}, 205149 (2016).
\bibitem{steffens2013} P. Steffens, C. H. Lee, N. Qureshi, K. Kihou, A. Iyo, H. Eisaki, and M. Braden, Phys. Rev. Lett. {\bf 110}, 137001 (2013).
\bibitem{zhang2014} C. Zhang,Y. Song, L.-P. Regnault, Y. Su, M. Enderle, J. Kulda, G. Tan, Z. C. Sims, T. Egami, Q. Si, and P. Dai, Phys. Rev. B {\bf 90}, 140502(R) (2014).
\bibitem{wwang2017}	W. Wang, J. T. Park, R. Yu, Y. Li, Y. Song, Z. Zhang, A. Ivanov, J. Kulda, and P. Dai, Phys. Rev. B {\bf 95}, 094519 (2017).
\bibitem{babkevich} P. Babkevich, B. Roessli, S. N. Gvasaliya, L.-P. Regnault, P. G. Freeman, E. Pomjakushina, K. Conder, and A. T. Boothroyd, Phys. Rev. B {\bf 83}, 180506(R) (2011).
\bibitem{mwma2017} M. Ma, P. Bourges, Y. Sidis, Y. Xu, S. Li, B. Hu, J. Li, F. Wang, and Y. Li, Phys. Rev. X {\bf 7}, 021025 (2017).
\bibitem{ysong2016} Y. Song, H. Man, R. Zhang, X. Lu, C. Zhang, M. Wang, G. Tan, L.-P. Regnault, Y. Su, J. Kang, R. M. Fernandes, and P. Dai, Phys. Rev. B {\bf 94}, 214516 (2016).
\bibitem{yhgu} Y. Gu, Z. Liu, T. Xie, W. Zhang, D. Gong, D. Hu, X. Ma, C. Li, L. Zhao, L. Lin, Z. Xu, G. Tan, G. Chen, Z. Y. Meng, Y. F. Yang, H. Luo, and S. Li, Phys. Rev. Lett. {\bf 119}, 157001(2017).
\bibitem{ckang2017} C. -J. Kang, T. Birol, and G. Kotliar, Phys. Rev. B {\bf 95}, 014511 (2017).
\bibitem{xing2016} X. Xing, W. Zhou, N. Zhou, F. Yuan, Y. Pan, H. Zhao, X. Xu, and Z. Shi, Supercond. Sci. Technol. {\bf 29}, 055005 (2016).


\end{thebibliography}

\begin{thebibliography}{}
\bibitem{txie} T. Xie, D. Gong, W. Zhang, Y. Gu, Z. Huesges, D. Chen, Y. Liu, L. Hao, S. Meng, Z. Lu, S. Li, and H. Luo, Supercond. Sci. Technol. {\bf 30}, 095002 (2017).


\end{thebibliography}
\end{document}